\documentclass[useAMS,usenatbib]{mnras}
\usepackage{graphicx}
\usepackage{epstopdf,color,verbatim}
\usepackage{threeparttable}
\usepackage{amsmath}
\usepackage{amssymb}
\usepackage[]{txfonts}

\def\simless{\mathbin{\lower 3pt\hbox
{$\rlap{\raise 5pt\hbox{$\char'074$}}\mathchar"7218$}}}   
\def\simmore{\mathbin{\lower 3pt\hbox
{$\rlap{\raise 5pt\hbox{$\char'076$}}\mathchar"7218$}}}   
\newcommand{\be}{\begin{equation}}
\newcommand{\ee}{\end{equation}}
\newcommand{\bi}{\begin{itemize}}
\newcommand{\ei}{\end{itemize}}

\newcommand{\unit}[1]{\,{\rm #1}}

\newcommand{\eq}[1]{Eq.~(\ref{eq:#1})}
\newcommand{\comp}{c/\omega_{\rm p}}

\newcommand{\ompt}{\omega_{\rm p}t}

\newcommand{\rhot}{r_{\rm L,hot}}
\newcommand{\gammarad}{\gamma_{\rm rad}}
\newcommand{\gammacool}{\gamma_{\rm cool}}
\newcommand{\gammasig}{\gamma_{\sigma}}

\def\bi{\begin{itemize}}
\def\ei{\end{itemize}}
\def\bq{\begin{equation}}
\def\eq{\end{equation}}
\def\bqy{\begin{eqnarray}}
\def\eqy{\end{eqnarray}}

\usepackage{epsfig} 
\usepackage{epstopdf}

\defcitealias{sironi_spitkovsky_2011}{SS11}
\defcitealias{cortes_sironi_2022}{CS22}
\defcitealias{cortes_sironi_2024}{CS24}

\title[]
{Particle Acceleration and Nonthermal Emission at the Intrabinary Shock of Spider Pulsars. II: Fast-Cooling Simulations}

\author[J. Cort\'es \& L. Sironi]
{Jorge Cort\'{e}s$^{1}$\thanks{E-mail:
\href{mailto:jicortes@ucsc.edu}{jicortes@ucsc.edu}} \& Lorenzo Sironi$^{2,3}$\thanks{E-mail:
\href{mailto:lsironi@astro.columbia.edu}{lsironi@astro.columbia.edu}}\\
$^{1}$Department of Astronomy and Astrophysics, University of California, Santa Cruz, CA 95064, USA\\
$^{2}$Department of Astronomy, Columbia University, 550 West 120th street, New York, NY, 10027, USA\\
$^{3}$Center for Computational Astrophysics, Flatiron Institute, 162 5th avenue, New York, NY, 10010, USA
}

\begin{document}
\label{firstpage}
\date{Received / Accepted}
\pagerange{\pageref{firstpage}--\pageref{lastpage}} \pubyear{2024}

\maketitle

\begin{abstract}
Spider pulsars are binary systems composed of a millisecond pulsar and a low-mass companion. Their X-ray emission, varying with orbital phase, originates from synchrotron radiation produced by high-energy electrons accelerated at the intrabinary shock. For fast-spinning pulsars in compact binary systems, the intrabinary shock emission occurs in the fast cooling regime. 
Using global two-dimensional particle-in-cell simulations, we investigate the effect of synchrotron losses on the shock structure and the resulting emission, assuming that the pulsar wind is stronger than the companion wind (so, the shock wraps around the companion), as expected in black widows. We find that the shock opening angle gets narrower for greater losses; the lightcurve shows a more prominent double-peaked signature (with two peaks  just before and after the pulsar eclipse) for stronger cooling; below the cooling frequency, the synchrotron spectrum displays a hard power-law range, consistent with X-ray observations.
\end{abstract}

\begin{keywords}
acceleration of particles --- magnetic reconnection 
--- radiation mechanisms: non-thermal --- shock waves
\end{keywords}

\section{Introduction}
Spider pulsars are compact binary systems composed of a millisecond pulsar and a low-mass companion---either a non-degenerate star with mass $\sim 0.2-0.4\,M_{\odot}$ (redbacks), or a degenerate star with mass $\ll 0.1\, M_{\odot}$ (black widows). The relativistic, magnetically-dominated pulsar wind interacts with the companion wind or magnetosphere, leading to the formation of an intrabinary shock (IBS) \citep{phinney_1988, fruchter_1990, arons_tavani_1993}. The IBS is a site of efficient particle acceleration and nonthermal emission \citep{harding_gaisser_1990,arons_tavani_1993}, primarily observed in the X-ray band \citep{huang_2012, bogdanov_2014,bogdanov_21,roberts_2015}. The X-ray emission from spider pulsars often exhibits orbital modulation \citep{bogdanov_2015, wadiasingh_2017, kandel_romani_an_2019}, providing insights into the geometry of the IBS and the physics of particle acceleration. The observed X-ray flux typically has two peaks, which has been attributed to Doppler effects caused by
the fast post-shock flow \citep{romani_sanchez_2016,sanchez_romani_2017,wadiasingh_2017,wadiasingh_2018,kandel_romani_an_2019,kandel_21,vandermerwe_2020}.

Observations of spider pulsars challenge conventional models of particle acceleration in relativistic shocks. X-ray spectra are markedly hard, with photon indices $\Gamma_X \sim 1 - 1.5$ in black widows and even harder in redbacks \citep{cheung_2012, romani_2014, arumugasamy_2015,kandel_romani_an_2019,swihart_2022,sullivan_romani_24}, implying electron energy distributions $dN/d\gamma \propto \gamma^{-p}$ with power-law indices $p=2\,\Gamma_X-1 \sim 1 - 2$. Such hard spectra are inconsistent with the standard theory of first-order Fermi acceleration at relativistic shocks \citep{fermi_49}, which yields steeper spectra with $p>2$ \citep[for a review, see][]{sironi_15}. This discrepancy suggests that alternative acceleration mechanisms, such as magnetic reconnection, may play a significant role. 

Near the pulsar equatorial plane, the pulsar wind consists of toroidal magnetic field stripes with alternating polarity, separated by current sheets \citep{bogo_99, petri_lyubarsky_2007}. At the IBS, the stripes are compressed and the oppositely-directed fields annihilate via shock-driven reconnection. Fully-kinetic particle-in-cell (PIC) simulations---zooming in near the IBS apex---have demonstrated that shock-driven reconnection produces power-law particle spectra with a slope as hard as $p=1$ \citep{sironi_spitkovsky_2011, lu_2021}. However, the {\it local} approach adopted by these studies does not allow to capture the {\it global} IBS dynamics, which is typically investigated with fluid-type simulations \citep{bogo_08,bogo_12,bogo_19,boschramon_2012,boschramon_barkov_perucho_2015,lamberts_2013,huber_2021}.

In recent years, {\it global} fully-kinetic PIC simulations of pulsars in binary systems---whose companion is either a pulsar or a normal star---have become possible \citep{cortes_sironi_2022,cortes_sironi_2024,romei_24,zhong_24}. In spider pulsars, global scales (i.e., the shock curvature radius $R_{\rm curv}$) are just three orders of magnitude greater than microscopic plasma scales (i.e., the typical post-shock Larmor radius $\rhot$), well within the reach of modern PIC simulations. In fact, the ratio of $R_{\rm curv}$ to the wavelength of the striped wind $\lambda=2\pi c/\Omega$ (here, $\Omega$ is the pulsar spin frequency) is
\begin{equation}
\label{eq:rcurv_lam}
    \frac{R_{\rm curv}}{\lambda}\sim 5\times 10^1\left ( \frac{R_{\rm curv}}{10^{10}\unit{cm}} \right ) \left ( \frac{\Omega}{10^3\unit{s^{-1}}} \right ).
\end{equation}
The ratio of stripe wavelength to the typical post-shock Larmor radius $\rhot$ is \citep{sironi_spitkovsky_2011}
\begin{equation} \label{eq:lam_comp}
    \frac{\lambda}{\rhot} \sim 4\times 10^1 \left ( \frac{\kappa}{10^4} \right ) \left ( \frac{10^{11}\,\mathrm{cm}}{{d_{\rm IBS}}} \right ) \left ( \frac{10^3\,\mathrm{s^{-1}}}{\Omega} \right )
\end{equation}
assuming a wind multiplicity \citep{goldreich_julian_1969} of $\kappa \sim 10^4$ \citep{harding_11,timokhin_harding_2015} and a distance between the shock and the pulsar of $ d_{\rm IBS}\sim 10^{11}\rm cm$ (\citealt{cortes_sironi_2022,cortes_sironi_2024}; respectively, \citetalias{cortes_sironi_2022,cortes_sironi_2024}). A multiplicity of $\kappa \sim 10^4$ may be an overestimate for millisecond pulsars \citep{harding_11}, so the ratio in Eq.~\ref{eq:lam_comp}  might be closer to unity.

In our earlier papers \citepalias{cortes_sironi_2022,cortes_sironi_2024}, we neglected electron cooling losses. For efficient magnetic dissipation via shock-driven reconnection, 
the typical Lorentz factor of post-shock electrons is $\gamma_\sigma =\omega_{\rm LC}/2 \Omega \kappa$, where $\omega_{\rm LC}=e B_{\rm LC}/m c$ is the electron Larmor frequency at the light cylinder radius $R_{\rm LC}=c/\Omega$. Electron cooling losses need to be included if the system is fast-cooling, i.e., if the cooling time of $\gamma\sim \gamma_\sigma$ electrons is shorter than the dynamical time 
\begin{equation}\label{eq:dyn}
t_{\rm dyn}=\frac{R_{\rm curv}}{c}\sim 0.3\,\left ( \frac{R_{\rm curv}}{10^{10}\unit{cm}} \right )\,{\rm s}
\end{equation}
where we have assumed that the post-shock flow is nearly relativistic. Most of the emission is expected to come from distances $\sim 10\, R_{\rm curv}$, so the estimate in Eq.~\ref{eq:dyn} is a conservative lower limit.

The synchrotron cooling time can be computed by extrapolating the field from the light cylinder to the IBS as $B= B_{\rm LC}(R_{\rm LC}/d_{\rm IBS})$, as appropriate for a toroidal field. The field at the light cylinder is related to the dipolar field strength $B_{\rm P}$ at the surface as $B_{\rm LC} \sim B_{\rm P} (R_{\rm NS} / R_{\rm LC})^3$, where $R_{\rm NS} \simeq 10\,\mathrm{km}$ is the neutron star radius. The cooling time  at the characteristic Lorentz factor $\gamma\sim \gamma_\sigma$ is then
\begin{equation}
     t_{\rm cool} \sim1.9\times 10^2  \left ( \frac{10^3\,\mathrm{s^{-1}}}{\Omega} \right )^6 \left ( \frac{10^9\,\mathrm{G}}{B_{\rm P}} \right )^3 \left ( \frac{d_{\rm IBS}}{10^{11}\,\mathrm{cm}} \right )^2 \left ( \frac{\kappa}{10^4} \right )\,\,\mathrm{s}
\end{equation}
The strong dependence of $t_{\rm cool}$ on pulsar and orbital parameters implies that compact systems harboring a fast-spinning millisecond pulsar will likely be in the fast-cooling regime, with $t_{\rm cool}\lesssim t_{\rm dyn}$. The same holds if the multiplicity is much smaller than $\kappa \sim 10^4$.

In this work, we employ global two-dimensional particle-in-cell simulations and investigate the effect of synchrotron cooling losses on the IBS structure and the resulting emission, assuming that the pulsar wind is stronger than the companion wind (so, the shock wraps around the companion), as expected in black widows. This work then extends our earlier papers \citepalias{cortes_sironi_2022,cortes_sironi_2024} to the fast-cooling regime. We find that: (\textit{i}) the shock opening angle gets narrower for greater cooling losses; 
(\textit{ii}) when the pulsar spin axis is nearly aligned with the orbital angular momentum, the light curve displays two peaks, just before and after the pulsar eclipse; the peaks get more pronounced for stronger cooling; (\textit{iii}) below the cooling frequency, the spectrum displays a hard power-law range; for strong cooling, the spectral peak reaches the synchrotron burnoff limit \citep{dejager_1992}. 

The paper is organized as follows. We describe our simulation setup in Section \ref{sec:sims}. We present our results in Section \ref{sec:results}, showing how the flow structure, the particle and synchrotron spectra, and the synchrotron lightcurves depend on the strength of cooling losses. We conclude in  Section \ref{sec:conclusion} and discuss the implications of our findings.

\section{Simulation Setup}
\label{sec:sims}
We use the 3D electromagnetic PIC code \textsc{TRISTAN-MP} \citep{buneman_1993, spitkovsky_2005}. We employ a 2D spatial domain in the $x-y$ plane, but we track all three components of velocity, electric current, and electromagnetic fields. Aside from the inclusion of synchrotron cooling, our setup parallels very closely what we employed in \citetalias{cortes_sironi_2022}, which we repeat here for completeness.

Since the distance between the pulsar and the intrabinary shock is $d_{\rm IBS}\gtrsim R_{\rm curv}$, we assume that the pulsar wind can be modeled as a sequence of plane-parallel stripes. The magnetically-dominated electron-positron pulsar wind propagates along $-\hat{x}$. It is injected from a moving boundary, that starts just to the right of the companion and moves along $+\hat{x}$ at the speed of light $c$. An absorbing layer for particles and fields is placed at $x=0$ (leftmost boundary). Periodic boundaries are used along  $y$. The magnetic field in the pulsar wind is initialized as
\begin{equation}
    B_y(x,t) = B_0\,\mathrm{tanh} \left\{ \frac{1}{\Delta} \left[ \alpha + \mathrm{cos} \left( \frac{2\pi(x+\beta_0ct}{\lambda} \right) \right] \right\}
    \label{eq:BB}
\end{equation}
where $\beta_0=(1-1/\gamma_0^2)^{1/2}$ is the  wind velocity and $\gamma_0=3$ the bulk Lorentz factor (a higher $\gamma_0$ yields identical results, apart from an overall shift in energy scale). The magnetic field flips across current sheets of hot plasma, having a thickness $\sim \Delta\lambda$. The field strength $B_0$ is parametrized via the magnetization $\sigma \equiv B_0^2/4 \pi \gamma_0 m n_{\rm 0} c^2=10$ (i.e., the ratio of Poynting to kinetic energy flux). Here, $m$ is the electron (or positron) mass and $n_{\rm 0}$ the density of particles in the ``cold wind'' (i.e. the region outside of current sheets). Finally, $\alpha$ quantifies the field averaged over one wavelength, such that $\langle B_y\rangle_\lambda/B_0=\alpha/(2-|\alpha|)$. We employ a value of $\alpha=0$---or equivalently, ``positive'' and ``negative'' stripes of comparable width---appropriate for the equatorial plane of the pulsar wind.

The relativistic skin depth in the cold wind $\comp \equiv (\gamma_0 m c^2 / 4 \pi e^2 n_{\rm 0})^{1/2}$ is resolved with 10 cells, where $e$ is the positron charge. It follows that the pre-shock Larmor radius $r_{\rm L} \equiv \gamma_0 m c^2 / e B_0=(\comp)/\sqrt{\sigma}$ is resolved with 3 cells, for $\sigma=10$. The post-shock Larmor radius, assuming complete field dissipation, is $\rhot=\sigma r_{\rm L}=(\gamma_0\sigma) mc^2/eB_0\simeq \gamma_\sigma mc^2/eB_0$, where we defined $\gamma_\sigma=\gamma_0(1+\sigma)$ as the mean particle Lorentz factor assuming full dissipation. For $\comp=10$ cells and $\sigma=10$, $\rhot$ is resolved with $\sim 30$ cells. The numerical speed of light is 0.45 cells/timestep. Within the cold wind, each computational cell is initialized with two pairs of cold ($kT/mc^2 = 10^{-4}$) electrons and positrons. The temperature in the current sheets is set by pressure balance, which yields a thermal spread $kT_{h}/mc^2=\Theta_h = \sigma / 2\eta$, where we choose that current sheets are denser than the striped wind by a factor of $\eta=3$.

Our computational domain is $4800\,\comp$ wide in the $y$ direction. The center of the companion is placed at $(x_c,y_c)=(1500,2400)\,\comp$, with a companion radius of $R_{\ast}=70\,\comp$. The companion surface (a cylinder, for our 2D geometry) is a conducting boundary for fields and a reflecting boundary for particles. The value for $R_{\ast}$ is chosen such that the companion wind (see below) is stopped by the pulsar wind at $R_{\rm curv} \simeq 200\,\comp$, which then gives the characteristic shock curvature radius. We set the stripe wavelength to be $\lambda=100\,\comp$, so that the ratio $\lambda/\rhot\simeq 30$, as expected in realistic systems. We then have $R_{\rm curv}/\lambda\simeq 2$, i.e., smaller than realistic cases by an order of magnitude. In \citetalias{cortes_sironi_2024}, we showed that our results are essentially the same for a larger companion, having $R_{\rm curv}/\lambda\simeq 4$. 

In our setup, the pulsar wind is stopped by a companion wind launched isotropically from its surface. We initialize an unmagnetized companion wind with realistic values of the radial momentum flux (twice larger than the momentum flux of the pulsar wind), but with artificially smaller particle density (and so, artificially higher wind velocity) to make the problem computationally tractable, and focus our computing efforts on pulsar wind particles. In the remainder of this work we will only consider acceleration and emission of pulsar wind particles.

\begin{figure*}
    \includegraphics[width=\textwidth, angle=0]{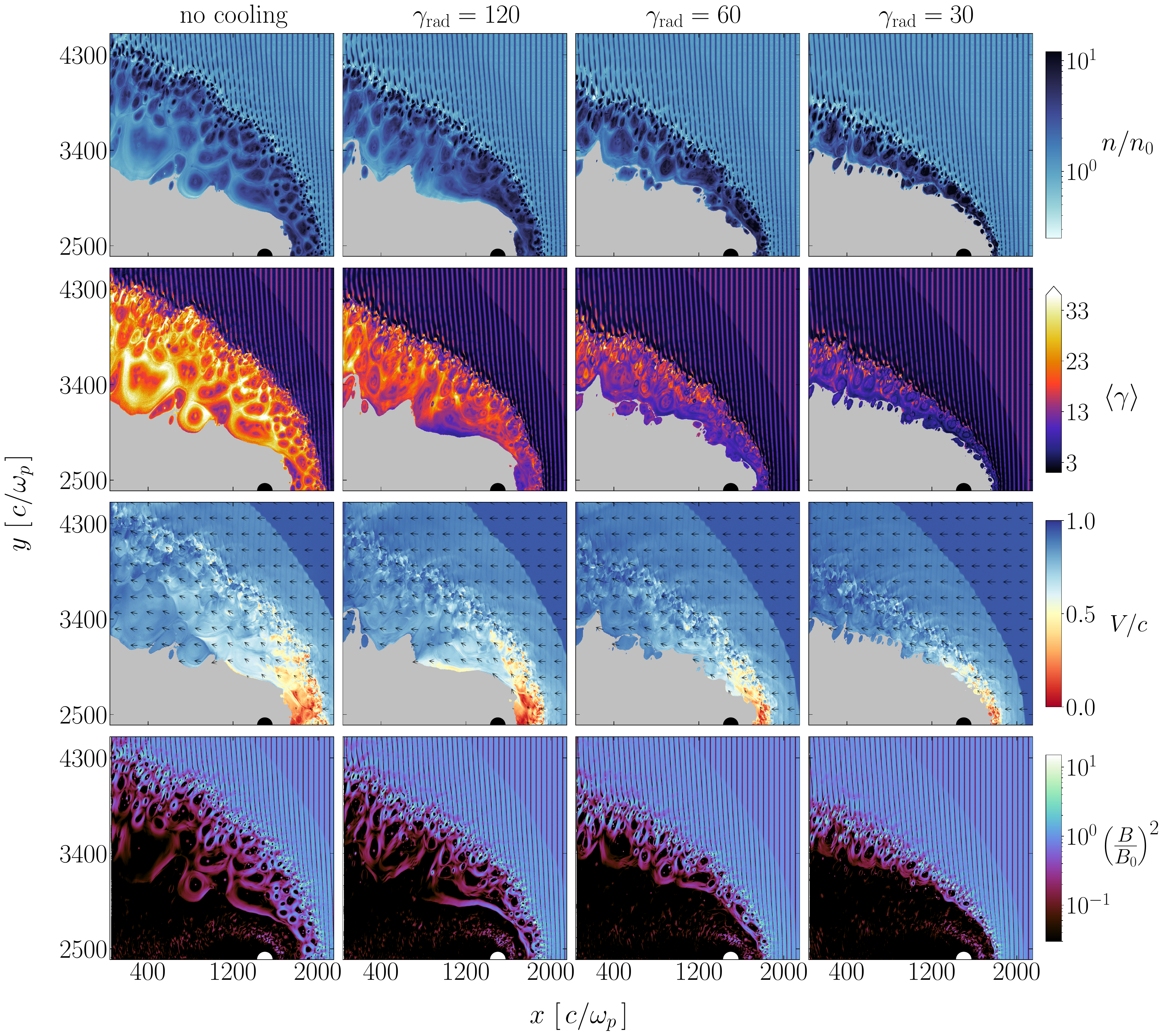}
    \caption{2D plots of a portion of the upper half of the domain for the uncooled case and the radiatively cooled cases with $\gamma_{\rm rad} = 120, 60$ and $30$ (left to right columns). We fix $\sigma = 10$ and the companion radius $R_\ast = 70\,\comp$ and we show results at $\ompt = 4050$. From top to bottom row, we show: \textbf{(1)} number density of pulsar wind particles in units of $n_0$; \textbf{(2)} mean particle Lorentz factor; \textbf{(3)} flow velocity of the pulsar wind in units of $c$, with arrows of unit length depicting flow direction; \textbf{(4)} magnetic energy density in units of the upstream value $B_0^2 / 8\pi$. Either a black circle (top three rows) or a white circle (bottom row) represents the companion star. In the top three rows, the grey region around and to the left of the companion
star is populated by companion wind particles, which we exclude from our analysis.}
    \label{fig:flow_4x4}
\end{figure*}

The change of momentum due to radiative losses is implemented using the reduced Landau-Lifshitz formalism \citep[e.g.,][]{vranic_2016}.
Given the importance of shock-driven reconnection in particle acceleration at the IBS \citepalias{cortes_sironi_2022,cortes_sironi_2024}, we characterize the strength of synchrotron cooling losses by defining $\gamma_{\rm rad}$, the classical radiation-reaction or ``burnoff'' limit \citep{dejager_1992}, at which the synchrotron radiation-reaction  drag force balances the accelerating force from the reconnection electric field $E_{\rm rec}$:
\begin{equation} \label{eq:RF}
    eE_{\rm rec} = \frac{4}{3}\sigma_{\rm T} \gamma_{\rm rad}^2 \frac{B_0^2}{8\pi}~
\end{equation}
where $\sigma_{\rm T}$ is the Thomson cross section.
In reconnection, $E_{\rm rec}$ is related to the reconnecting magnetic field $B_0$ as $E_{\rm rec}=\eta_{\rm rec}B_0$, where the so-called reconnection rate $\eta_{\rm rec}\simeq 0.1$ in the relativistic ($\sigma\gg1$) regime appropriate for pulsar winds \citep[e.g.,][]{kagan_15}. We investigate a range of cooling strengths, by varying $\gammarad$ from 30 to 120, or equivalently $\gammarad/\gammasig$ from 1 to 4 (moderately strong to weak cooling). We also compare our results to the uncooled case $\gammarad=\infty$. 
From our definition of $\gammarad$, one can find the Lorentz factor $\gammacool$ at which the synchrotron cooling time is comparable to the dynamical time $t_{\rm dyn}$. We find 
\begin{equation} \label{eq:gammacool}
\frac{\gammacool}{\gammasig}=\eta_{\rm rec}^{-1} \left(\frac{\rhot}{R_{\rm curv}}\right)\left(\frac{\gammarad}{\gammasig}\right)^2~~,
\end{equation}
where fast cooling is identified as $\gammacool\lesssim \gammasig$. In our simulations, $R_{\rm curv}/\rhot\simeq 63$. For $\eta_{\rm rec}\simeq 0.1$, the fast cooling regime is realized in our simulations for $\gammarad/\gammasig\lesssim 2.5$, so for $\gammarad=30$ and $\gammarad=60$.

As compared to \citetalias{cortes_sironi_2022,cortes_sironi_2024}, we employ thicker current sheets, with $\Delta \lambda / 2\pi = 5\,\comp$ rather than $\Delta \lambda / 2\pi = 2\,\comp$. If we were to use $\Delta \lambda / 2\pi = 2\,\comp$ also for cases with strong cooling ($\gammarad=30$ and 60),  reconnection would be initiated well ahead of the IBS. In order to isolate the effect of shock-driven reconnection (i.e., reconnection should only be due to interaction with the IBS), we therefore choose to increase the sheet thickness as compared to our earlier works. We have checked that, for the uncooled case, the results are the same between runs with $\Delta \lambda / 2\pi=2\,\comp$ and $\Delta \lambda / 2\pi=5\,\comp$.

\section{Results}
\label{sec:results}
In this section we present our results. First we describe how radiative cooling losses affect the flow dynamics; then we show the dependence of the synchrotron emission signatures (spectrum, emissivity, lightcurve) on $\gammarad$; finally we comment on the role of $\gammarad$ in inhibiting particle acceleration to $\gamma\gg\gammasig$. 

\subsection{Flow Dynamics}
\label{flow}
Figure \ref{fig:flow_4x4} illustrates the global morphology of the four simulations presented in this work. They range from the uncooled case ($\gamma_{\rm rad} = \infty$) in the leftmost column to the strongest cooled case ($\gammarad=30$) in the rightmost column. From top to bottom, the rows display: the number density of pulsar wind particles; the mean particle Lorentz factor; the flow velocity of the pulsar wind;  the magnetic energy density. All panels refer to $\ompt = 4050$, when the shock has reached a quasi-steady state, as quantified in \citetalias{cortes_sironi_2024}. Either a black circle (top three rows) or a white circle (bottom row) represents the companion star. In the top three rows, the grey region around and to the left of the companion
star is populated by companion wind particles, which we exclude from our analysis.

While the general flow characteristics---such as the formation of the IBS, the presence of the upstream fast magnetohydrodynamic (MHD) shock, and the development of plasmoids via shock-driven magnetic reconnection---remain consistent with those presented in the uncooled cases of \citetalias{cortes_sironi_2022, cortes_sironi_2024}, strong synchrotron cooling losses introduce notable differences, which we now describe. 

Perhaps the most drastic effect, in regards to the structure of the IBS and the post-shock region, is in the area covered by the shocked pulsar wind, which substantially shrinks with increasing cooling losses. As $\gammarad$ decreases, the mean particle Lorentz factor drops (second row), which reduces the post-shock pressure. The ram pressure of the pre-shock pulsar wind, which is insensitive to $\gammarad$, then confines more effectively the downstream flow at lower $\gammarad$, as compared to the uncooled case, causing the shock opening angle to shrink. In addition to this macroscopic/global effect, cooling losses also affect the force balance in individual plasmoids. 
Previous PIC simulations of collisionless relativistic reconnection under the influence of synchrotron losses \citep[e.g.,][]{hayk_2019} have shown that greater cooling makes the plasmoids more compressible, leading to smaller sizes and larger central overdensities. This trend is apparent in the post-shock region of our simulations.

The mean post-shock  Lorentz factor of pulsar wind particles, shown in the second row of Figure \ref{fig:flow_4x4}, is systematically lower for stronger cooling. In the uncooled case (left column), $\langle \gamma \rangle$ peaks on the outskirts of plasmoids, regardless of whether they are near the shock or farther downstream. In contrast, for strong cooling ($\gammarad=30$, right column), $\langle \gamma \rangle$ is greatest near the shock (especially near its high-latitude wings), and drops farther downstream, indicating that active heating/acceleration is mostly localized in the shock vicinity. In the strongest cooled case, the region hosting hot particles appears to move away from the apex of the shock faster than the corresponding uncooled case (compare left and right panels in the third row, in the downstream region with $x\lesssim 400\,c/\omega_{\rm p}$). As we demonstrate below, this has an effect on the observed synchrotron lightcurves.
 
\subsection{Particle Energy Spectra and Synchrotron Spectra}
\label{sync}

\begin{figure}
    \includegraphics[width=\columnwidth, angle=0]{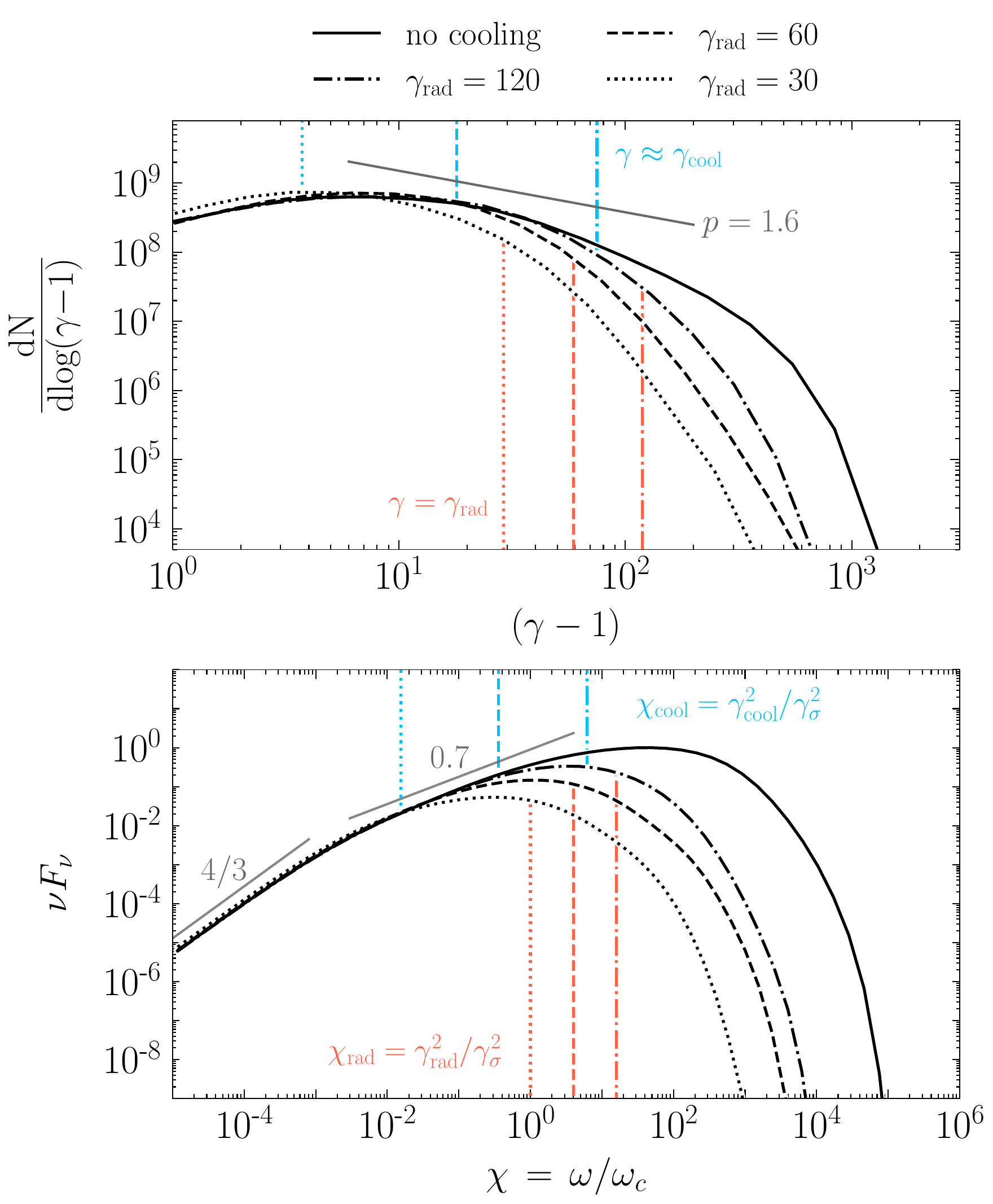}
    \caption{\textbf{Top: } Downstream particle energy spectra for the uncooled case (solid) and cooled runs with $\gamma_{\rm rad} = 120$ (dot-dashed), $60$ (dashed), and $30$ (dotted), evaluated at $\ompt=4050$ (the same time as Figure \ref{fig:flow_4x4}). Vertical lines, matching the style of the respective spectra, indicate $\gamma_{\rm rad}$ (lower, red) and $\gamma_{\rm cool}$ (upper, blue), as obtained from Eq. \ref{eq:gammacool}. For comparison, the average post-shock Lorentz factor $\gamma_\sigma=\gamma_0(1+\sigma)\simeq 30$, assuming complete field dissipation. \textbf{Bottom: }Corresponding angle-integrated synchrotron spectra $\nu F_\nu$, with the same line styles. We define $\chi=\omega/\omega_{\rm c}$, where the characteristic synchrotron frequency $\omega_{\rm c}=\gamma_\sigma^2 eB_0/mc$ is calculated for $\gammasig$. Vertical lines, matching the style of the respective spectra, indicate $\chi_{\rm rad}=(\gammarad/\gammasig)^2$ (lower, red) and $\chi_{\rm cool}=(\gammacool/\gammasig)^2$ (upper, blue).}
    \label{fig:spectra}
\end{figure}
\begin{figure*}
    \includegraphics[width=\textwidth, angle=0]{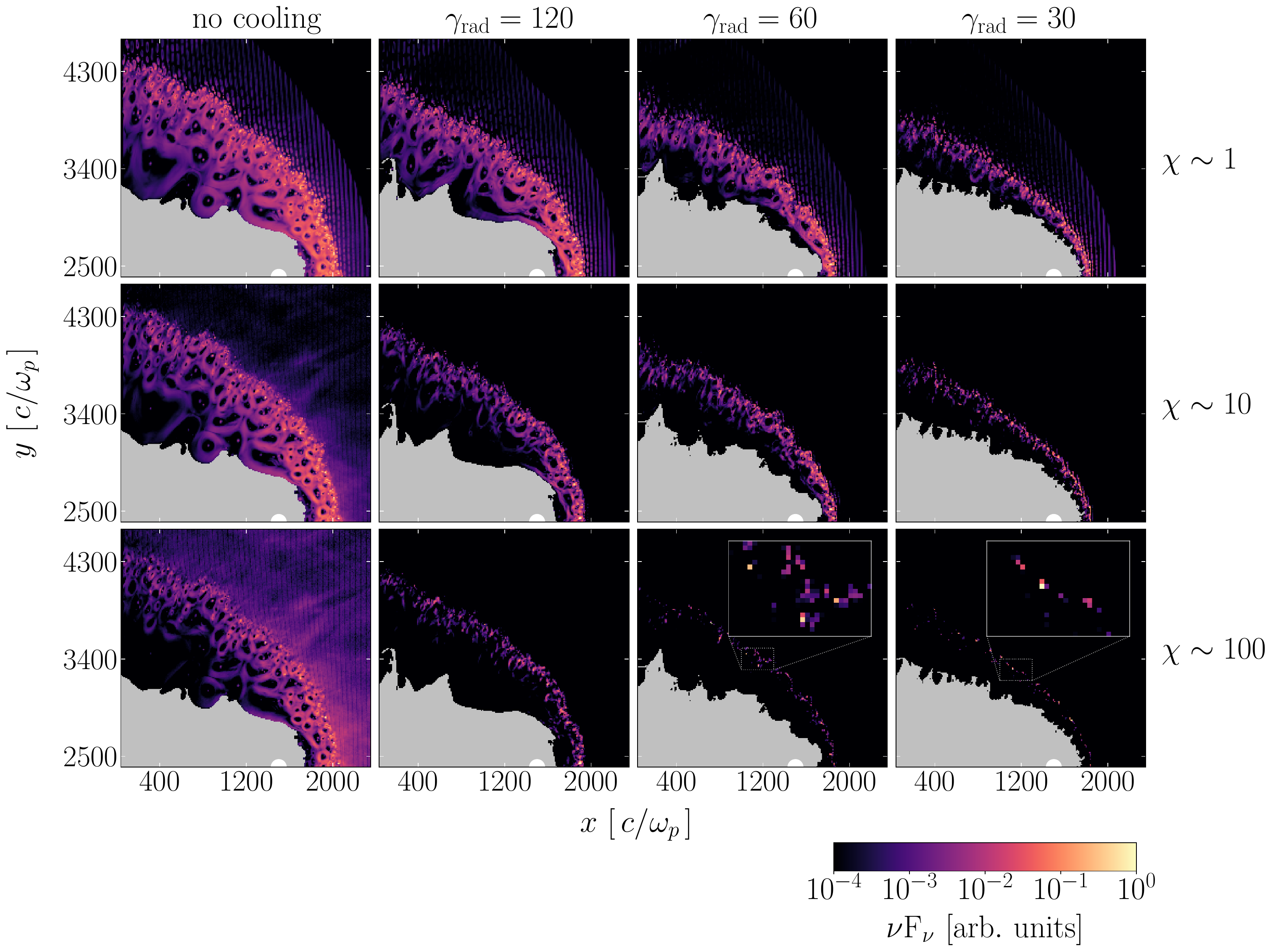}
    \caption{2D plots of the angle-integrated synchrotron emissivity at $\ompt=4050$. From left to right, columns correspond to the uncooled case and cooled runs with $\gamma_{\rm rad} = 120,\,60$, and $30$. Rows from top to bottom display the emissivity for different frequencies: $\chi=\omega/\omega_{\rm c}=1,\,10$, and $100$. The emissivity in each panel is normalized to the maximum value within that panel. A white circle represents the companion star. The grey region around and to the left of the companion star is populated by companion wind particles, which we exclude from our analysis.
    }
    \label{fig:sync_em_iso}
\end{figure*}

Particle energy spectra and synchrotron spectra are presented in Figure \ref{fig:spectra} at $\ompt=4050$. For both sets of spectra, only the contribution from pulsar wind particles in the post-shock flow (i.e., behind the IBS) is taken into account. The top panel of Figure \ref{fig:spectra} shows the downstream particle spectra $dN / d \ln{(\gamma-1)}$ for our four cases. Vertical lines, matching the style of the respective spectra, indicate $\gamma_{\rm rad}$ (lower, red) and $\gamma_{\rm cool}$ (upper, blue), as obtained from Eq. \ref{eq:gammacool}.

In the absence of cooling (solid line), the particle spectrum can be described as a broad distribution. Within the range from $\gamma \sim \gamma_0 = 3$ (the initial bulk flow Lorentz factor) to $\gamma \sim \gamma_\sigma \sim 30$ (the Lorentz factor achieved by particles in the case of complete field dissipation), the spectrum can be described by a power law, $dN/d\gamma \propto (\gamma - 1)^{-p}$, with a hard slope of $p \simeq 1.6$. In the cooled cases, the particle spectrum below the cooling break ($\gamma\lesssim \gammacool$) is the same as the uncooled spectrum, while cooled spectra above the cooling break are steeper than the uncooled case. For $\gammarad=30$ and 60, we find that $\gammacool\lesssim \gammasig$, i.e., the system is in the fast-cooling regime, whereas the simulation with $\gammarad=120$ is in the slow-cooling regime. 

Regardless of $\gammarad$, we find that some particles can be accelerated beyond the nominal synchrotron burnoff limit, i.e., up to $\gamma\gtrsim \gammarad$. Local PIC simulations of reconnection have demonstrated that in the strong cooling regime electrons can accelerate beyond the standard burnoff Lorentz factor in regions where the magnetic field component perpendicular to the particle momentum is weak, which suppresses synchrotron losses \citep{cerutti_13,cerutti_14a,chernoglazov_23}. 
Our global simulations of spider pulsars yield similar conclusions, and they show that the fraction of particles exceeding $\gammarad$ is greater for stronger cooling (i.e., smaller $\gammarad$).

This trend carries over to the synchrotron spectra, shown in the bottom panel of Figure \ref{fig:spectra}. They are calculated following \citet{kirk_reville_2010}, by summing over the angle-integrated synchrotron emission from every particle in the downstream region. The synchrotron frequency on the horizontal axis is normalized to the characteristic frequency $\omega_{\rm c} = \gamma_\sigma ^2 e B_0 / mc$ emitted by particles with $\gamma = \gamma_\sigma$. For all spectra, the expected $\nu F_\nu \propto \nu^{4/3}$ scaling is observed at low frequencies. Beyond this, at $\chi=\omega/\omega_{\rm c} \gtrsim 10^{-2}$, the uncooled spectrum transitions to a scaling $\nu F_\nu \propto \nu^{0.7}$ up to $\chi \sim 10$. The slope of the uncooled synchrotron spectrum in this range follows from the power-law slope $p\simeq 1.6$ of the particle energy spectrum. Above the cooling break, i.e., for  $\chi\gtrsim \chi_{\rm cool}=(\gammacool/\gammasig)^2$ (vertical blue lines), the cooled spectra fall below the corresponding uncooled case. The peak of the synchrotron spectrum recedes to lower frequencies for increasing cooling strength; yet, in all cases a substantial fraction of the synchrotron power is emitted at frequencies exceeding the nominal burnoff limit $\chi_{\rm rad}=(\gammarad/\gammasig)^2$ (vertical red lines).

Figure \ref{fig:sync_em_iso} presents the angle-integrated synchrotron emissivity, for different cooling strengths (columns) and different frequencies (rows). The figure shows that, for all cooled cases, most of the emission comes from the downstream region (which we have indeed used to calculate  the spectra of Figure \ref{fig:spectra}). The same holds for the uncooled case (left column), apart from the highest frequency ($\chi=100$), where high-energy particles streaming back upstream from the shock \citepalias{cortes_sironi_2022,cortes_sironi_2024} appreciably contribute to the synchrotron flux.
For strong cooling (and, at fixed cooling, for higher $\chi$), the synchrotron emissivity is sharply concentrated in the near downstream region, i.e., just behind the IBS. This parallels closely the trend observed in the mean particle Lorentz factor (second row of Figure \ref{fig:flow_4x4}). The synchrotron flux at high frequencies ($\chi=100$) for small $\gammarad$ is dominated by few bright regions, likely hosting recent episodes of efficient reconnection-driven acceleration.

\begin{figure}
    \begin{center}
        \includegraphics[width=\columnwidth, angle=0]{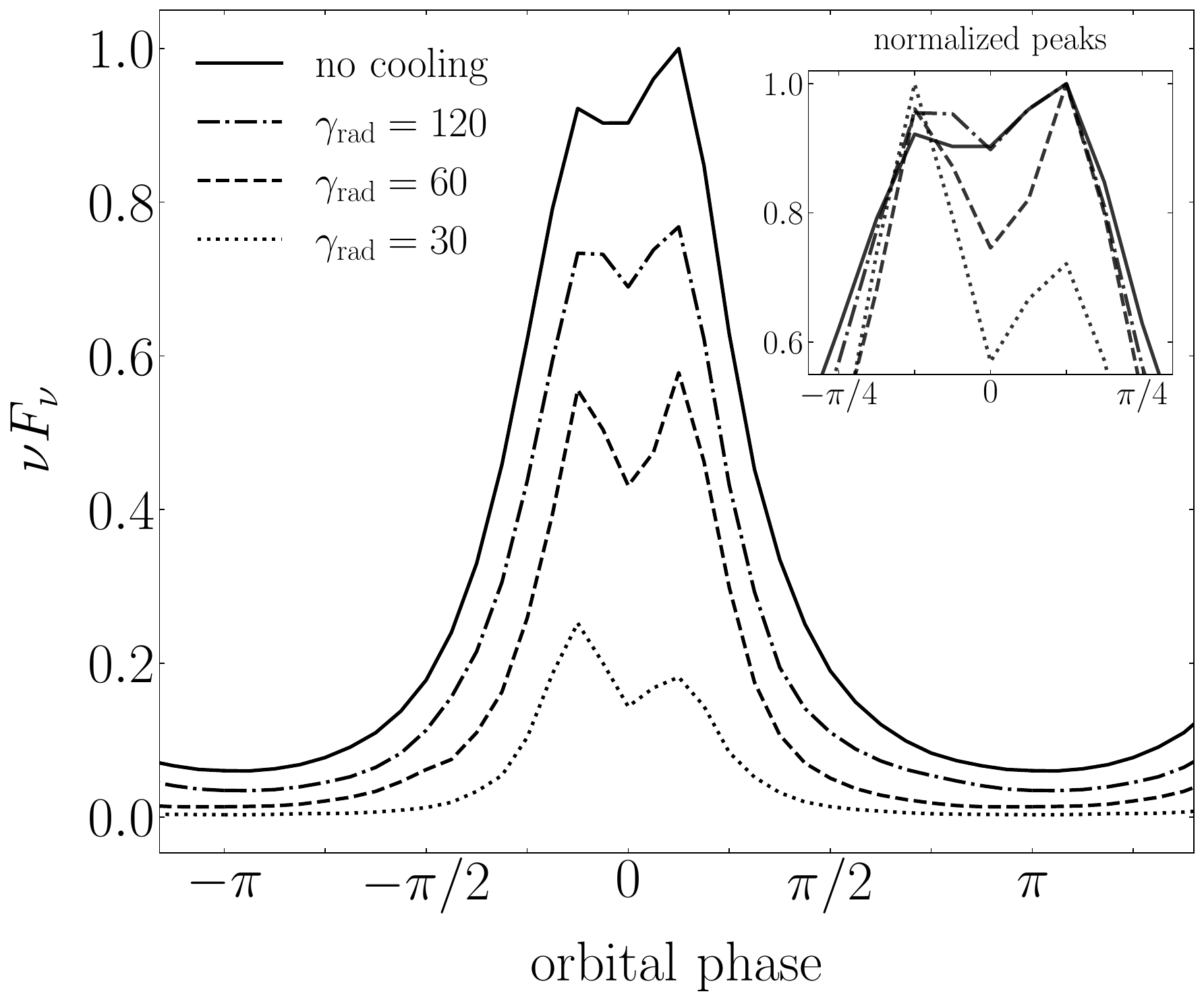}
        \caption{Phase-resolved synchrotron lightcurves at $\chi\equiv\omega/\omega_{\rm c}=1$ for the uncooled case (solid) and cooled runs with $\gamma_{\rm rad}=120$ (dot-dashed), $60$ (dashed), and $30$ (dotted), evaluated at $\ompt=4050$. The orbital phase $\phi$ is defined such that $\phi=0$ corresponds to superior conjunction (when the pulsar is eclipsed), while $\phi=\pm\pi$ marks inferior conjunction. The main panel shows lightcurves normalized to the peak value of the uncooled case, whereas the inset---sharing the same axes as the main panel----displays the same curves normalized to their respective peak, highlighting differences in the double-peaked feature.}
        \label{fig:syncLC}
    \end{center}
\end{figure}

\begin{figure*}
    \includegraphics[width=\textwidth, angle=0]{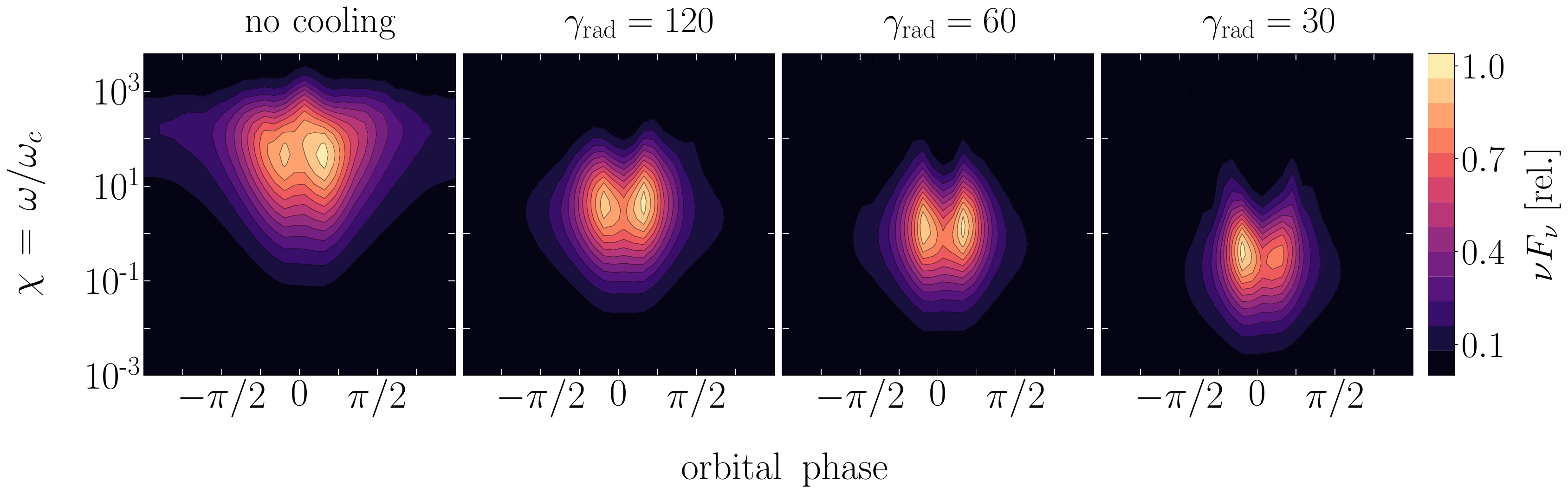}
    \caption{Contour plots of the $\nu F_\nu$ intensity as a function of orbital phase $\phi$ (horizontal axis) and  normalized frequency $\chi=\omega/\omega_{\rm c}$ (vertical axis), for different levels of cooling. Each panel is taken at $\ompt=4050$ and normalized to its peak value. Only the contribution from pulsar wind particles in the post-shock flow (i.e., behind the IBS) is taken into account.   
    }
    \label{fig:syncLC_chi}
\end{figure*}

\begin{figure*}
    \includegraphics[width=\textwidth, angle=0]{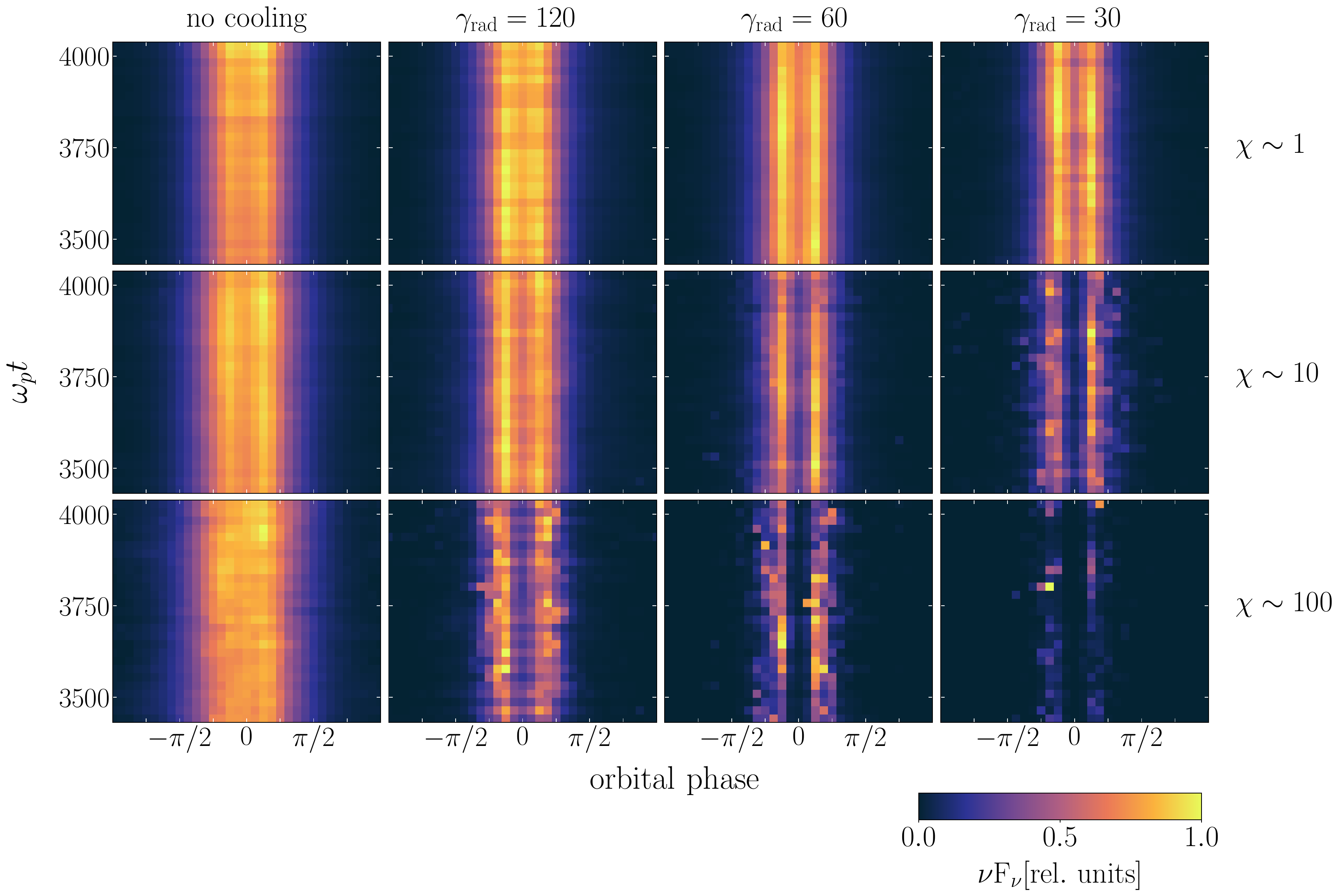}
    \caption{Contour plots of the $\nu F_\nu$ intensity as a function of orbital phase $\phi$ (horizontal axis) and  time $\ompt=3442 - 4050$ (vertical axis). From left to right, columns correspond to the uncooled case and cooled runs with $\gamma_{\rm rad} = 120,\,60$, and $30$; rows, from top to bottom, refer to normalized synchrotron frequencies of $\chi=1,\,10$, and $100$. Each panel is normalized to its peak value. Only the contribution from pulsar wind particles in the post-shock flow (i.e., behind the IBS) is taken into account.
    }
    \label{fig:LCs_time}
\end{figure*}

\subsection{Synchrotron Lightcurves}
\label{emission}

Phase-resolved lightcurves at $\ompt=4050$ are presented in  Figure \ref{fig:syncLC} for different levels of cooling, at a fixed frequency $\omega=\omega_{\rm c}$. In the main panel, all curves are normalized to the peak value of the uncooled case (solid line), while the inset highlights relative differences by normalizing the curves to their respective peak value. The characteristic double-peaked structure, located at orbital phases $\phi \simeq \pm \pi/8$ (just before and after superior conjunction), is present across all cases, regardless of the cooling strength. However, the fractional drop from the peaks (at $\phi \simeq \pm \pi/8$) to the trough (at $\phi=0$) is greater for stronger cooling: in the $\gamma_{\rm rad}=30$ case, the flux in the trough is smaller than the peak flux by $40\%$, compared to a drop of only $10\%$ in the uncooled case.

Figures \ref{fig:syncLC_chi} and \ref{fig:LCs_time} show the dependence of the lightcurve on frequency and time, respectively. In both figures, only the contribution from pulsar wind particles in the post-shock flow (i.e., behind
the IBS) is taken into account. Figure \ref{fig:syncLC_chi} shows that the double-peaked feature generally persists across a broad range of frequencies, even though the peak frequency shifts to lower values with increasing cooling strength, as already demonstrated in Figure \ref{fig:spectra}. The uncooled case shows a double-peaked lightcurve up to $\chi\sim 10^2$, but at higher frequencies the lightcurve has a single peak at superior conjunction (i.e., $\phi=0$). In contrast, in cooled cases the double-peaked signature appears at all frequencies above the peak frequency. Furthermore, for $\gamma_{\rm rad}=30$, the phase separation between the two peaks increases at higher frequencies, and the peaks get sharper (i.e., the ratio of peak-to-trough intensity is greater). For strong cooling, high-energy particles---dominating the high-frequency part of the synchrotron spectrum---are confined to a narrow strip just downstream of the IBS (see Figure \ref{fig:flow_4x4} [second row] and Figure \ref{fig:sync_em_iso}). There, the flow bulk speed is consistently tangential to the shock surface, i.e., nearly uni-directional (Figure \ref{fig:flow_4x4} [third row]). This enhances Doppler boosting effects \citep{romani_sanchez_2016,sanchez_romani_2017,wadiasingh_2017,wadiasingh_2018,kandel_romani_an_2019,kandel_21,vandermerwe_2020}, causing a more pronounced double-peaked signature for stronger cooling and higher frequencies. 

Figure \ref{fig:LCs_time} assesses the temporal variability of the lightcurves at different frequencies. We confirm the trends seen in  Figure \ref{fig:syncLC_chi}: for stronger cooling, the phase separation between the two peaks increases at high frequencies; also, the ratio of peak-to-trough intensity is greater, especially at high frequencies. Figure \ref{fig:LCs_time} confirms that the double-peaked signature is robust at all times, regardless of the level of cooling and the observed frequency (with the exception of $\chi=100$ in the uncooled case). The two peaks are generally comparable in intensity, although significant temporal variations exist, especially for strong cooling and high frequencies. We note that {\it systematic} asymmetries in X-ray lightcurves (with one peak brighter than the other) have been attributed to the orbital motion of the system \citep{romani_sanchez_2016, wadiasingh_2017}. Our simulations do not include the effect of orbital motion, so any temporal variation should be attributed to the {\it stochastic} nature of plasmoid mergers and ensuing particle acceleration in the post-shock flow.  

\begin{figure*}
    \begin{center}
        \includegraphics[width=0.9\textwidth, angle=0]{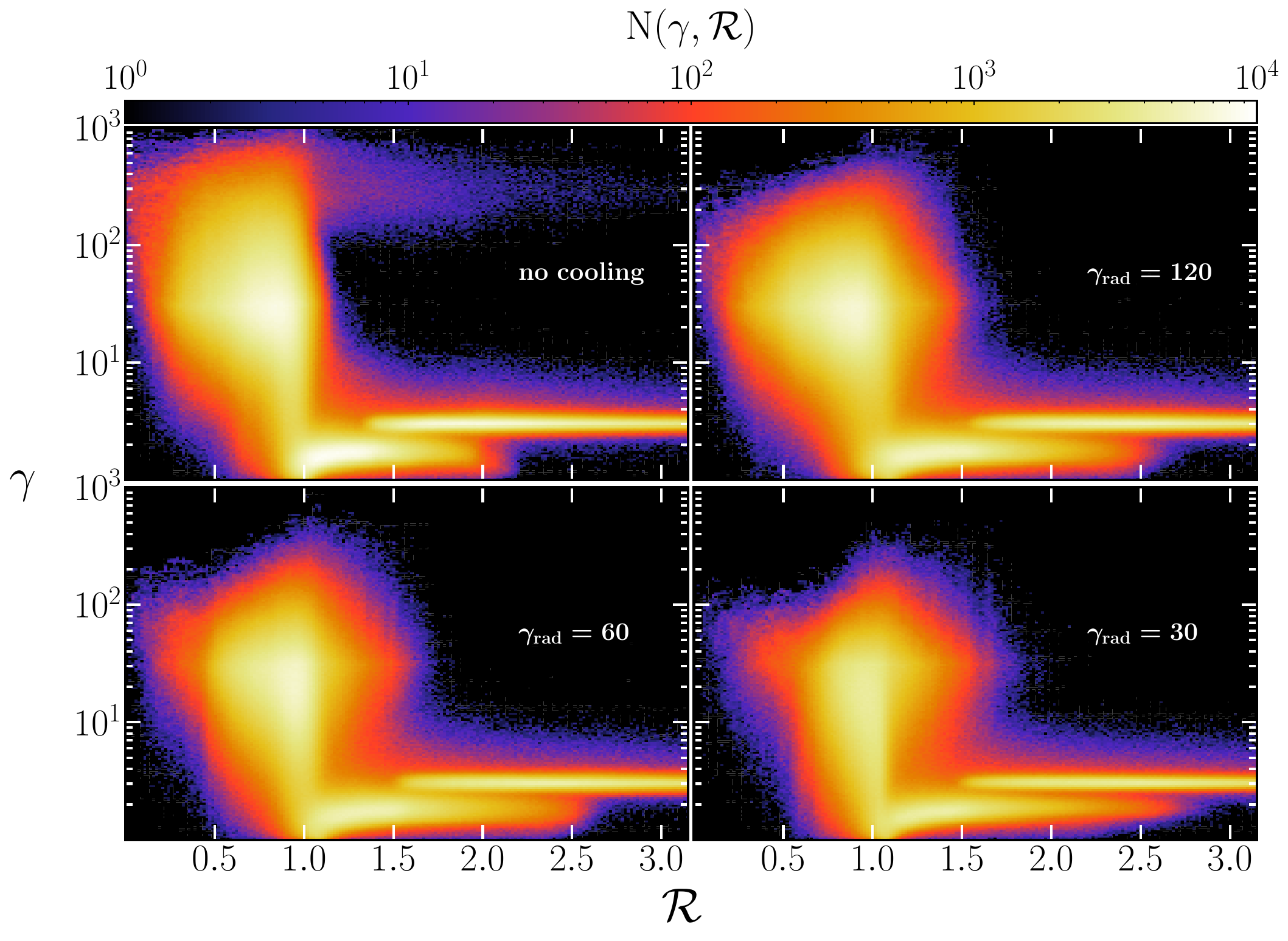}
        \caption{2D histograms of particles tracked in $\gamma - \mathcal{R}$ space, for different levels of cooling. All particles are selected such that their Lorentz factor is $\gamma \geq \gammasig\sim 30$ at some point within the time range $\ompt=2025-4050$. Each particle appears as many times as it is saved.
        }
        \label{fig:tau}
    \end{center}
\end{figure*}

\subsection{Particle Acceleration beyond $\gammasig$}
\label{accel} 

In \citetalias{cortes_sironi_2022, cortes_sironi_2024}, we demonstrated that shock-driven magnetic reconnection efficiently energizes the pulsar wind particles up to, and slightly beyond, a typical Lorentz factor $\gamma \sim \gamma_\sigma = \gamma_0(1+\sigma)$. This process remains robust in the presence of synchrotron cooling losses. However, acceleration to higher energies ($\gamma \gg \gamma_\sigma$) is significantly affected by the level of cooling, as we now describe.

In the uncooled case, particles energized by shock-driven reconnection can propagate back into the upstream if their Lorentz factor is $\gamma/\gammasig\gtrsim (\lambda/\rhot)/4\pi\sim 2.5$. In order to propagate far ahead of the shock, a particle must be able to traverse half of the stripe wavelength (and so, reverse its sense of gyration in the upstream alternating field) before being overtaken by the shock. In the upstream, the particle is then accelerated by a mechanism akin to the pick-up process widely discussed in space physics \citep[e.g.,][]{mobius_85,iwamoto_2022}: the particle is accelerated by the motional electric field while gyrating around the upstream field. Cooling losses will inhibit the propagation of high-energy particles upstream of the shock, if the cooling rate is greater than the pick-up acceleration rate. 

We quantify this effect in Figure \ref{fig:tau}. For different levels of cooling, we track the particles whose Lorentz factor exceeds $\gammasig\sim 30$ at some point within the time range $\ompt=2025-4050$. We then plot 2D histograms of the tracked particles in the $\gamma - \mathcal{R}$ plane.
The quantity $\mathcal{R}$ is defined as follows. At each time, we fit the shape of the IBS with an ellipse, whose centre is at $(x_0,y_0=y_c)$. At each time, the best-fitting values for the semi-major ($a_p$) and semi-minor ($b_p$) axes are then used to compute
\begin{equation}
\mathcal{R} = \frac{\left ( x-x_0 \right )^2}{a_p^2} + \frac{\left ( y-y_0 \right )^2}{b_p^2}
\end{equation}
for a particle having coordinates $(x,y)$. It follows that $\mathcal{R}$ serves as a proxy for the particle position relative to the IBS. Based on $\mathcal{R}$, a particle can be classified as residing in: (i) downstream, $\mathcal{R} \lesssim 0.9$; (ii) IBS, $0.9 \lesssim \mathcal{R} \lesssim 1.1$; or (iii) upstream, $\mathcal{R} \gtrsim 1.1$.

The figure shows that some features are common to all cases. The pulsar wind comes towards the shock with a typical Lorentz factor $\gamma \simeq \gamma_0 = 3$. The flow is then slightly decelerated at the fast MHD shock ($\mathcal{R} \sim 2-2.5$), where its typical Lorentz factor decreases down to $\gamma \simeq 2$. Upon interaction with the IBS ($\mathcal{R} \sim 1$), rapid energization by shock-driven reconnection pushes particles up to $\gamma \sim \gamma_\sigma \simeq 30$. Eventually, most of the particles reside in the downstream ($\mathcal{R} \lesssim 0.9$).

The main difference between the uncooled case and all cooled cases---including the $\gammarad=120$ case of weakest cooling---is the fact that cooled cases lack high-energy particles ($\gamma\gtrsim 10^2$) residing in the upstream ($\mathcal{R} \gtrsim 1.1$). Even a moderate level of synchrotron cooling is sufficient to inhibit particle acceleration due to the pick-up process described in \citetalias{cortes_sironi_2024}. For all $\gammarad\lesssim 120$, high-energy particles are confined solely to the post-shock flow. Even weaker cooling, i.e. $\gammarad\gtrsim 120$, would be required to establish how cooled cases transition to the uncooled case, as regard to the physics of upstream pick-up acceleration. 
This can be understood by comparing a few important timescales. In order to propagate far ahead of the shock, a particle must be able to traverse half of the stripe wavelength (and so, reverse its sense of gyration in the upstream alternating $B_{y}$ field) before being overtaken by the shock. This criterion is best phrased in the upstream frame, where the time to cross half of the stripe wavelength is $\gamma_0 \lambda/2c$. For a relativistic shock seen in the upstream frame, particles returning upstream are caught up by the shock after completing a fraction $\sim \gamma_0^{-1}$ of their Larmor orbit, and so after a time $\gamma_0^{-1}(2\pi/\omega_{\rm L})$, where $\omega_{\rm L}=e B_0/\gamma_0^2 \gamma mc$ is the upstream Larmor frequency for a particle having Lorentz factor $\gamma$ in the downstream frame. 
At the same time, in the presence of cooling losses, the synchrotron cooling time in the upstream field needs to be longer than the gyration time (more precisely, a fraction $\sim \gamma_0^{-1}$ of the gyration time). The two conditions can be written (see also \citetalias{cortes_sironi_2024}) as
\begin{equation}
\frac{1}{4\pi}\frac{\lambda}{\rhot}\ll \frac{\gamma}{\gammasig}\ll \left(\frac{1}{2\pi \,\eta_{\rm rec}}\right)^{1/2} \left(\frac{\gammarad}{\gammasig}\right)
\end{equation}
and they can be both be satisfied only if 
\begin{equation}
\frac{\gammarad}{\gammasig}\gg  \left(\frac{\eta_{\rm rec}}{8\pi}\right)^{1/2} \frac{\lambda}{\rhot}~.
\end{equation}
For $\eta_{\rm rec}\simeq 0.1$ and our choice of $\lambda/\rhot\simeq 30$, this requires $\gammarad/\gammasig\gg2$, which is not satisfied by our simulations (our largest value of the radiation reaction Lorentz factor is $\gammarad=4\,\gammasig$).

\section{Summary and Discussion}
\label{sec:conclusion}
We have employed global 2D particle-in-cell simulations and investigated the effect of synchrotron cooling losses on the IBS structure and the resulting emission, assuming that the shock wraps around the companion star, as expected in black widows. Global kinetic simulations allow to capture the shock dynamics concurrently with the physics of field dissipation and particle acceleration, thus overcoming the limitations of fluid-type simulations and semi-analytical models. This work extends our earlier papers \citepalias{cortes_sironi_2022,cortes_sironi_2024} to the fast-cooling regime, $\gammacool\lesssim \gammasig$. We find that: (\textit{i}) the shock opening angle gets narrower for greater cooling losses, due to the drop in post-shock plasma pressure; 
(\textit{ii}) when the pulsar spin axis is nearly aligned with the orbital angular momentum, the light curve displays two peaks, just before and after the pulsar eclipse; the peaks get more pronounced for stronger cooling; (\textit{iii}) below the cooling frequency, the synchrotron spectrum displays a hard power-law range; for strong cooling, the spectral peak reaches the synchrotron burnoff limit \citep{dejager_1992}. 

In our work, we have considered the role of cooling losses in the case that the radiation-reaction Lorentz factor $\gammarad\gtrsim \gammasig$ and the cooling Lorentz factor $\gammacool\lesssim \gammasig$ (in our runs, this is satisfied for $\gammarad=30$ and 60). For realistic spider systems, the characteristic Lorentz factor $\gammasig$ is
\begin{equation} \label{eq:gammasig}
\gamma_\sigma =\frac{e B_{\rm LC}}{2 m c \Omega \kappa}\simeq 3.3\times 10^4 \left ( \frac{\Omega }{10^3\,\mathrm{s^{-1}}} \right )^2 \left ( \frac{B_{\rm P}}{ 10^9\,\mathrm{G}} \right )  \left ( \frac{ 10^4}{\kappa} \right )
\end{equation}
On the other hand, the synchrotron burnoff Lorentz factor is 
\begin{equation}
    \label{eq:grad}
    \gamma_{\rm rad}=\sqrt{\frac{ 6\pi e \eta_{\rm rec}}{\sigma_{\rm T}B_{\rm IBS}}}\simeq 1.1\times 10^7\! \left ( \frac{ 10^3\,\mathrm{s^{-1}} }{\Omega} \right ) \left ( \frac{10^9\,\mathrm{G} }{ B_{\rm P}} \right )^{1/2}  \left ( \frac{\eta_{\rm rec}}{0.1} \right )^{1/2} \left ( \frac{d_{\rm IBS}}{10^{11}\,{\rm cm}} \right )^{1/2}
\end{equation}
where $B_{\rm IBS}=B_{\rm P}(R_{\rm NS}/R_{\rm LC})^3(R_{\rm LC}/d_{\rm IBS})$ is the field at the IBS. It follows that in realistic spider systems $\gammarad\gg \gammasig$, whereas our simulations have  $\gammarad$ just above $\gammasig$. In order to preserve the fast-cooling regime $\gammacool\lesssim \gammasig$, and at the same time allow for realistic $\gammarad\gg \gammasig$, one would need simulations with much larger $R_{\rm curv}/\rhot$ than we have employed here, see Equation \ref{eq:gammacool}. In future work, we will fix $\gammacool/\gammasig$ and increase $\gammarad/\gammasig\propto (R_{\rm curv}/\rhot)^{1/2}$, to validate the robustness of our conclusions towards the realistic regime $\gammarad\gg \gammasig$. 

Our simulations employ a magnetization of $\sigma=10$, lower than in realistic pulsar winds. In the uncooled case, \citetalias{cortes_sironi_2024} showed that higher magnetizations lead to similar results (in terms of shock dynamics, synchrotron spectrum and lightcurve), aside from an overall shift in energy scale (as encoded in $\gammasig$), and a moderate trend for harder X-ray spectra at higher magnetizations (with $\nu F_\nu\propto \nu$ in the limit $\sigma\gg1$). We argue that radiatively-cooled simulations with $\sigma\gg10$ would lead to similar results as in this work, as long as the ratios $\gammarad/\gammasig$, $\gammacool/\gammasig$ and $R_{\rm curv}/\rhot$ are preserved (see Equation \ref{eq:gammacool}). Further work is needed to validate this conjecture.

We conclude with a few caveats, as we have already discussed in the non-radiative cases of \citetalias{cortes_sironi_2024}. First, we have employed 2D simulations, and we defer to future work an assessment of 3D effects, which can alter the IBS geometry as well as the physics of particle acceleration in reconnection \citep[e.g.,][]{zhang_sironi_21,zhang_sironi_23}. 3D simulations will also be able to provide a first-principles assessment of the synchrotron polarization properties as a function of frequency and orbital phase \citep{sullivan_23}. Second, we have neglected the orbital motion of the system, which has been invoked to explain asymmetries in the light curve \citep{kandel_21}. Third, the thickness of the pre-shock current sheets is chosen such that reconnection does not spontaneously start before the shock; in reality, magnetic field dissipation should start already since the pulsar light cylinder \citep{cerutti_philippov_2017,cerutti_philippov_dubus_2020}. Finally, we have assumed that the pulsar wind can be modeled as a sequence of plane-parallel stripes. This is appropriate if $d_{\rm IBS}\gg R_{\rm curv}$, whereas for realistic spider systems $d_{\rm IBS}$ is not much larger than $R_{\rm curv}$. In a realistic 3D configuration, this also implies that one cannot assume a single value of $\alpha$, since $\alpha$ depends on latitude \citepalias{cortes_sironi_2024}.

\section*{Acknowledgements}
We thank A. Sullivan for useful discussions. J.C. acknowledges support provided by the NSF MPS-Ascend Postdoctoral Research Fellowship under grant no. AST-2402292. L.S. acknowledges support from DoE Early Career Award DE-SC0023015, NASA ATP 80NSSC24K1238, NASA ATP 80NSSC24K1826, and NSF AST-2307202. This work was supported by a grant from the Simons Foundation (MP-SCMPS-00001470) and facilitated by the Multimessenger Plasma Physics Center (MPPC), grant PHY-2206609. Computational resources were  provided by Columbia University (Ginsburg) and by facilities supported by the Scientific Computing Core at the Flatiron Institute, a division of the Simons Foundation.

\section*{Data Availability}
The simulated data underlying this paper will be shared upon reasonable request to the corresponding author(s).

\bibliographystyle{mnras}
\bibliography{spiders_master,spider2,araa}

\label{lastpage}
\end{document}